# Gobi: WebAssembly as a Practical Path to Library Sandboxing

Shravan Narayan  
UC San Diego

Tal Garfinkel  
Stanford University

Sorin Lerner  
UC San Diego

Hovav Shacham  
UT Austin

Deian Stefan  
UC San Diego

## Abstract

Software based fault isolation (SFI) is a powerful approach to reducing the impact of security vulnerabilities in large and critical C/C++ applications like Firefox and Apache. Unfortunately, practical SFI tools have not been broadly available.

Developing an SFI toolchain is a significant engineering challenge. Only in recent years have browser vendors invested in building production quality SFI tools NaCl/P-NaCl to sandbox mobile code. Further, without committed support, these tools are not viable, e.g.,NaCl has recently been discontinued, orphaning projects that relied on it.

WebAssembly (Wasm) offers a promising solution—it can support high performance sandboxing—and has been embraced by all major browser vendors—thus seems to have a viable future. However, Wasm presently only offers a solution for sandboxing mobile code. Providing SFI for native application, such as C/C++ plugins or libraries requires additional steps.

To reconcile the different worlds of Wasm on the browser and native platforms, we present Gobi. Gobi is a system of compiler changes and runtime support that can sandbox normal C/C++ libraries with Wasm—allowing them to be compiled and linked into native applications. Gobi has been tested on libjpeg, libpng, and zlib.

Based on our experience developing Gobi we offer a detailed analysis of how SFI developers can address the difficult challenge of maintaining an SFI toolchain long term. We conclude with a call to arms to the Wasm community and SFI research community to make Wasm based module sandboxing a first class use case and describe how this can significantly benefit both communities.

**Addendum:** This short paper was originally written in January of 2019. Since then, the implementation and design of Gobi evolved substantially and some of the issues raised in this paper have since been addressed by the Wasm community. Some challenges still remain. We thus left the paper largely intact and only provide a brief update on the state of WebAssembly tooling as of November 2019 in the last section.

## 1 Introduction

Software-fault isolation (SFI), introduced in 1993 [29], is an effective technique for reducing trust in software components in both userland and kernels [28]. But the dominant mechanism for module isolation today is process-based privilege separation and sandboxing [6,23], not SFI. Process-based isolation carries with it IPC overhead that can be substantial for tightly coupled modules. In this paper, we argue that WebAssembly, developed for accelerating Web content, is a viable basis for general-purpose SFI that is performant and maintainable.

Unlike process-based isolation, SFI does not enjoy instruction-set support on widely deployed processors. Accordingly, SFI vendors need to develop and maintain their own compiler backend (to rewrite modules to restrict memory accesses), their own linker and loader (to install modules in the host address space), and their own base library and ABI (to facilitate interaction between the module and the outside world). This is a large undertaking. Google's Native Client (NaCl) SFI effort was largely rewritten in the port from 32-bit x86 [30] to 64-bit x86 [26], and the most recent version of LLVM for which Google has supplied NaCl backend patches is 3.7.0, released in late 2015.

WebAssembly (Wasm) [14] has supplanted NaCl for high-performance Web content sandboxing in browsers. It is a viable basis for general purpose SFI. Wasm already provides the core SFI guarantee: isolated modules can access memory only within a single region. Unlike NaCl, Wasm was designed for hosting isolated modules in-process, rather than in a dedicated NaCl process. And Wasm's popularity—it is implemented in all major browsers [18]—means that it is supported by a robust and growing toolchain ecosystem.

We have built a prototype Wasm SFI system, called Gobi. Our experience highlights several shortcomings of the current Wasm ecosystem for applications beyond Web content acceleration. Some shortcomings are already being addressed by the community. For example, browsers use their existing JIT infrastructure to compile Wasm bytecode to machine code. For host processes that don't already include a JIT, a standalone machine code generator is needed—and is currently being developed as Cranelift [8]. Others are not currently being addressed. For example, the C/C++ ABI implemented by Wasm compilers and browser hosts has not been standardized or even fully documented. We sound a call-to-action to the Wasm community to address these shortcomings.

Gobi is intentionally designed to minimize changes to the Wasm compiler, even at the cost of some performance. We are working to upstream our changes to the Emscripten compiler. We describe the design decisions we made to maximize the maintainability of Gobi within the Wasm ecosystem.

Preliminary compatibility and performance evaluation of Gobi supports the feasibility of Wasm as a base for general-purpose SFI. To date, Wasm performance has been compared against JavaScript performance, not native code. Perhaps as a result, the bounds-checking techniques employed by current Wasm runtimes are well behind the SFI state of the art. We thus sound a call-to-action to SFI researchers to adopt Wasm bytecode as an SFI intermediate representation and to bring high-performance sandboxing to Wasm.

## 2 Prototype

Our positions in this paper stem from our experience building Gobi. Therefore we briefly discuss the various requirements addressed by Gobi to leverage Wasm for module sandboxing.

By itself, Wasm provides an essential building block—memory isolation, i.e. a module can only access its own memory. However, we must safely use this memory isolated module in the application. In this section, we use an example of an image decoder application that wishes to use libjpeg (a JPEG decoding library) in a sandboxed manner. To setup, we compile an unmodified libjpeg with our custom Emscripten compiler to produce a Wasm module, which is in turn compiled with a Wasm compiler, such as Cranelift, to produce an ELF library that is used with the Gobi runtime.

**Runtime.** To run Wasm modules like libjpeg, we should only require a simple standardized Wasm runtime for which multiple open source implementations exist [3, 16, 25]. This runtime provides basic functions for Wasm module setup—providing APIs to extend or free memory for the Wasm module, mechanisms to correctly handle traps, implementation of data structures for indirect function call support (the Wasm specification requires all indirect calls to go through the runtime to ensure memory safety). In practice, the libjpeg Wasm module also requires the Emscripten specific "embedding" runtime. This runtime is providing support for C primitives and platform ABIs such as stack and heap separation in the address space, primitives for `setjmp` and `longjmp`, threading primitives, floating point primitives and clock/time primitives. Emscripten provides this runtime for Wasm as used on the web, it does not provide this runtime for non-web embeddings. We bridge this gap in Gobi.

**Syscalls.** The libjpeg Wasm module needs to invoke syscalls (e.g., for I/O or logging). However, Wasm only supports a small number of memory related syscalls in the Wasm runtime. Other syscalls are either unsupported or unofficially supported in the Emscripten embedding runtime. Gobi thus provide support for several syscalls, ensuring compatibility while retaining memory isolation.

**Module sandboxing APIs.** To safely use the sandboxed libjpeg module the Gobi runtime also provides several APIs to application code:

▶ **Function Calls.** The application calls into libjpeg to parse the JPEG image. However, the application and the libjpeg Wasm module have ABI differences. We provide an API that accounts for ABI differences (e.g., function name mangling, struct parameters, etc.).

▶ **Memory Allocation.** The application and libjpeg must share data such as the parameters of function calls. To allow controlled sharing, we provide an API that allocates memory in Wasm module memory. This memory is accessible to both the library and the application.

▶ **Pointer Swizzling.** Pointers in the Wasm libjpeg module are to be represented in the format specified in Wasm standard—indexes into the Wasm memory. We thus provide APIs to convert or "swizzle" pointers to the Wasm representation and its reverse.

▶ **Callback Registration.** The application may need the Wasm libjpeg module to invoke some application functions. For example, libjpeg needs to invoke an error handler application if it encounters a parsing error. We thus provide an API to register callbacks.

▶ **Threading.** To safely support multiple application threads calling into libjpeg simultaneously, we must create a new Wasm stack for each calling thread. We thus provide an API to create a new stack in the Wasm module which is called for each application thread.

**Compiler modifications.** For the application and Wasm libjpeg module to be compatible, they must have compatible machine models—identical sizes and alignments for datatypes such as an **int**, **long** and the size of a pointer. In practice these are not the same: the application uses the default machine model of the target platform, however, the Wasm libjpeg module uses the machine model dictated by the Wasm standard and toolchain. We resolve these differences by modifying the machine model used by Emscripten when compiling Wasm modules, taking care to minimize the amount of changed cod (see Section 3.1).

## 3 Challenges in SFI module sandboxing

In this section, we address what has been a historical challenge for SFI—maintaining an up to date runtime and compiler toolchain. Typically, SFI compiler toolchains are implemented as forks of existing compiler toolchains. These toolchains either require substantial engineering effort for maintenance or are simply unmaintained and lag behind when it comes to bug fixes, optimizations and language support. For instance, the NaCl compiler toolchain, when supported, required significant engineering effort to keep up with upstream changes, while research toolchains such (e.g., Memsentry [19]) are often not updated beyond their first release.

Below we show how various design choices in Gobi are explicitly made to keep it maintainable in the long term. These choices are not tied to the Wasm-based module sandboxing and may useful to other SFI compiler toolchains. We believe that SFI runtimes and compiler toolchains must similarly and explicitly make design choices that consider maintainability.

### 3.1 Maintaining Compiler modifications

As described in Section 2, machine model differences must be resolved in Emscripten to ensure compatibility of the application and sandboxed module code. Machine model changes, however, can lead to large changes in compiler implementations: several compiler phases (e.g., the code generation phase for all pointer instructions), optimizations and instructions (e.g., pointer instructions need to use different register classes) have to be adjusted to deal with the changed machine model. This is an expensive engineering effort. In our experience porting the machine model of the NaCl toolchain, this could take up to 4 months of engineering. Worse, the resulting toolchain changes are both hard to maintain and difficult to merge back upstream.

The alternate option of avoiding compiler modifications and transforming all data as it crosses between the application and the sandboxed module (and vice-versa) has its own drawbacks. In particular, this requires us to perform deep object copies as they cross the boundary. For languages like C and C++, this is not only slow, but is hard to do automatically as this process would require accurately tracking the underlying types of **void*** , sizes of C arrays etc. Works such as PtrSplit [20] optimize overhead for this by tracking this information only for potentially shared data; however, this still incurs overheads of about 13% and cannot be used for multi-threaded programs. We propose an alternative approach that that minimizes compiler changes without giving up on performance.

Our key insight is that most machine model differences can be resolved with trivial code modifications with the exception of pointer size differences. We can thus adjust machine model difference such as the size of **int** and **long** [1], the padding and alignment of datatypes, and struct field padding in the Wasm compiler toolchain and leave handling pointer size differences to the application.

To test the viability of this approach, we implement the proposed changes which requires a small patch of approximately 150 LOC to Emscripten's Clang frontend. Further, we also examine how we can efficiently handle pointer size difference at the application level below.

**Handling differences in pointer size.** Pointer size differences are resolved through a combination of application level data conversions and Emscripten modifications. To illustrate our solution, we discuss how the module sandboxing APIs(from Section 2) along with compiler modifications resolve differences for 3 pointer related datatypes.

*Pointers.* Wasm pointers are 4 bytes, while application pointers are 8 bytes on 64-bit platforms. We modify pointers to be the appropriate size as part of the swizzling APIs described in Section 2.

*Struct containing a pointer field.* To ensure structs containing pointers are compatible when crossing the application module boundary, we modify Clang to ensure pointer fields are aligned to 8 bytes and have a trailing padding of 4 bytes. This causes the field offsets of the structs in the Wasm module to be identical to the field offsets of the same struct in the application.

*Structs containing an array of pointers.* Structs with a field of type **void*[]** must be made compatible when crossing the application module boundary. Unfortunately, we cannot use the prior approach of adding trailing padding after a pointer element, the C ABI does not permit arrays to have padding between elements. We therefore use an alternate approach. First, to ensure that the structure sizes are the same for the application and the sandboxed module, we modify the Wasm toolchain to add trailing padding after the entire array equal to the size of the array. Second, we provide an API to the application that modifies the array *in place* to use the "sandbox" representation of 4-byte pointer elements when structs are copied from the application to the sandboxed module and vice-versa. [2] While this second step imposes a runtime cost, transferring arrays of pointers is relatively rare in practice which makes this an acceptable trade-off.

---

[1] Since different operating systems and architectures use different machine models, the user specifies the target machine model via a compiler flag.

[2] This same approach is used for parameters and returns of an array of pointers outside a struct

We note that while usage of such APIs may be inconvenient for developers, we believe it is possible to ensure such conversion APIs are automatically invoked in languages such as C++ which have strong meta-programming capabilities.

## 3.2 Maintaining syscall support

As described in Section 2, Gobi provides support for syscalls. Providing and maintaining safe syscall support requires substantial engineering effort. Syscalls must retain compatibility and functionality while ensuring memory isolation. A syscall specification in the Wasm standard would both improve Wasm compatibility and ease Wasm module sandboxing maintenance; the WebAssembly System Interface (WASI) [10] partly addresses this need (though was not available when we started our work on Gobi, see Section 7). We believe the SFI community should maintain a generic syscall support layer providing access to POSIX syscalls, that is adaptable to different SFI toolchains and is guaranteed not to break isolation. A good starting point here would be to use the existing implementation from the NaCl project and generalize it further. While, ongoing engineering effort would still be required to ensure maintenance for this code, we believe this is tractable: the syscall support may be restricted to commonly used syscalls—a list that does not change frequently. As a consequence of these challenges, wider support for syscalls in Gobi is currently an ongoing engineering effort.

## 4 Challenges in Wasm tooling

In the implementation of Gobi, we encountered several limitations in Wasm tooling that hindered our use of Wasm as an SFI toolchain. While some of these limitations are because the tools are relatively new, we believe others to be more fundamental and open problems. We describe the challenges and possible workarounds below.

**Generating binaries.** Tools such as Cranelift or Turbofan currently only support JIT compilation of Wasm modules, and cannot produce ELF binaries. We therefore use the wasm2c tool [3] which provides a one to one conversion of Wasm instructions to C statements. We compile the generated C code [3] using a C compiler to produce an ELF binary that is memory isolated. In the future, compiling Wasm modules in this circuitous manner can be avoided as Cranelift aims to support ahead of time compilation of Wasm modules.

**Runtime.** As described in Section 2, while we can leverage one of several maintained Wasm runtime implementations to allow module sandboxing, providing an Emscripten embedded runtime represents a fundamental weakness as the embedding runtime has not been standardized and is an adhoc set of functionality required by Emscripten. To address this challenge, Wasm would need to either standardize this so that the community may easily develop embedding runtimes, or alternately, Emscripten must maintain the non-web embedding runtime upstream. We note there are some new efforts to standardize the C-Wasm ABI [5] as well as some open source projects that aim to provide this non-web embedding runtime such as wasmer [4], however these are very nascent efforts and are missing large portions of the Emscripten runtime. As a consequence of this non-standardization the embedded runtime support in Gobi is only partial.

**No linking support.** Wasm sandboxed modules are currently dynamically loaded and do not support static or dynamic linking due to toolchain limitations. First, the toolchain generates common symbols for all Wasm modules causing name collisions when applications link multiple Wasm sandboxed modules. But, even if this problem is solved, per-session sandboxes (creating multiple sandboxes for the same module) would still result in name collisions. From our experience, we believe such challenges are more difficult to address than at first glance, and urge the Wasm community to explicitly take this up as part of Wasm tooling standard. Until then, we can easily provide APIs that simplify working with dynamically loaded modules and provide the same type-checking guarantees as static linking.

**Threading.** Gobi does not currently allow multiple threads simultaneously executing code in the Wasm module as we found Wasm thread support in Emscripten to be buggy. Wasm threading is brand new, and we believe these bugs will soon be resolved.

## 5 Is Wasm usable for module sandboxing?

To ensure there are no fundamental limitations in using Wasm for module sandboxing we evaluate three questions.

▶ **Compatibility.** Can we easily sandbox any library?

▶ **Correctness.** Do our tooling modifications or APIs cause incorrect behavior in modules?

▶ **Performance.** Is the overhead of Wasm module sandboxing reasonable?

## 5.1 Compatibility

To ensure compatibility of Wasm sandboxing we sandbox 3 commonly used open source libraries—libjpeg, libpng and zlib. First we modify the build scripts to use the custom Emscripten toolchain to output a Wasm module.

---

[3]The generated C code resembles assembly — this is in no way related to "decompilation" of the Wasm module.

We then use wasm2c and a standard C compiler to produce an ELF binary. Finally, we combine this with the Gobi runtime to produce the final library.

We find that while the Wasm toolchain retains a high compatibility with existing code, providing a complete embedding runtime implementation remains challenging as addressed in Section 4. Each new library we sandbox highlights different portions of the embedding runtime that we have missed and are forced to implement.

## 5.2 Correctness

To ensure the correctness of our Wasm sandboxing implementation, we must test that the libraries maintain correct behavior after 3 different stages—when they are compiled with our modified Wasm toolchain, when the libraries are used with the Gobi APIs and when the libraries and API are used in real world applications. First, to ensure our toolchain preserves libraries behavior, we ran the test suites for libjpeg, libpng and zlib all of which pass. Next, we created a test-suite for our APIs to ensure that both basic sandboxing functionality as well as edge case conditions such as data type overflow and memory isolation boundary addresses work correctly. Lastly, to ensure the libraries work in real world use cases, we modify the Firefox web browser to use these three Wasm sandboxed libraries, ensuring that Firefox produces identical web pages and images in all cases.

## 5.3 Performance

To understand the performance overhead of Wasm sandboxing, we measure the overhead of a simple application that decodes JPEG images and re-encodes the images at a different quality when using a sandboxed libjpeg. We also make some basic performance optimizations to the wasm2c toolchain eliminating bounds checks as described in the Wasm specification [14]. We measure the overhead to be 85%.

While the overhead here is high, we believe this is mostly due to the current state of Wasm tooling. For instance, wasm2c itself imposes large overhead and does not apply any SFI specific optimizations such as using a dedicated register to store the base address. Long term, the Wasm community is informally targeting reducing overhead to around 20% [2], putting it in the same ballpark as efficient SFI toolchains like NaCl. We believe applications may be willing to accept this overhead especially if the overhead is not on the critical path. For instance, we have tested that Firefox has no user visible slowdown on web-pages when using sandboxed image libraries having 20% overhead.

## 6 Empowering Wasm and SFI communities

In this section, we discuss how the Wasm and SFI research communities can work together and benefit from improving Wasm module sandboxing.

**Benefits to the Wasm community.** The Wasm community needs high performance memory isolation and the SFI research community has a long history of investigating hardware and software techniques for efficient isolation code on commodity hardware [12, 13, 19, 21, 24, 26, 30]. Many of these ideas could be applied to Wasm . Indeed some hardware based isolation techniques achieve overheads as low as 5%, much lower than Wasm's unofficial performance targets.

**Call to action for the Wasm community.** To facilitate SFI performance research on Wasm, the Wasm community must address two limitations that hinder its use as an SFI toolchain. First, the Wasm standard and tooling restrictions listed in Section 4, such as incompatible machine models and lack of embedding runtime standards must be addressed. Second, the Wasm standard must take great care during design so that they do not accidentally disallow performance optimizations.

To illustrate this, we give two examples. First, Wasm separates the stack and heap, i.e., the linear memory. Isolating the control transfer information—the return address—prevents classic stack smashing attacks. Unfortunately, it's not clear if separating local variables and function arguments meaningfully prevents data-only attacks [15]: most languages, when compiled to Wasm, will have to compile at least part of their stack to the untyped linear memory. (Richer types could have meaningful benefits though [11].) This design choice does, however, makes it more challenging to use hardware features such as Intel MPK [17] to efficiently enforce isolation; the current Wasm design, for example, would require switching between two separate keys—one for the linear memory and one for the stack.

Second, Wasm's CFI is similarly unnecessarily strong. Indirect jumps are only permitted if the target is a function with correct signature. NaCl, instead, optimizes CFI requirements by permitting jumps to any 32-byte aligned address and carefully padding instructions, trading better performance for larger binary sizes. Indeed, NaCl style CFI can be enforced purely in hardware in newer Intel CPUs using Intel CET [1]. The current Wasm standard has disallowed such trade-offs due to its CFI choice despite the fact that memory safety is not violated.

We believe the Wasm standard should embrace flexibility where possible, and allow implementations to explore different trade-offs. In the web scenario, this would allow browsers to choose between different Wasm implementations depending on the target platform's memory, operating system and hardware primitives.

**Benefits to the SFI community.** The SFI community can leverage Wasm for a task that has historically been very challenging—ensuring the availability of a well maintained SFI toolchain. It is likely that this would significantly increase the use of SFI in industry software, making SFI much more commonplace. A secondary benefit of a standard well maintained toolchain to the SFI research community is the ability to easily compare trade-offs between different SFI techniques in a uniform manner.

**Call to action for the SFI community.** Wasm is a place where SFI optimizations can have a huge impact on real world code. SFI researchers should ensure their solutions are compatible with Wasm so that the software industry benefits from their improvements to SFI. In leveraging existing platforms such as Wasm as an SFI toolchain, the SFI community must try to minimize the amount of changes to the existing toolchain so that the changes are maintainable. We show how this can be done in Section 3. Finally, the SFI community should view the Wasm standard as an intermediate representation (IR) for SFI and must actively participate in the Wasm standards process to ensure that the IR does not rule out specific SFI optimizations or implementations.

## 7 Updates as of November 2019

This paper was originally written in January of 2019. Over the last year, some of the issues raised in this paper have been addressed by the Wasm community. The most significant changes for Wasm based module sandboxing are the creation of (1) WASI (WebAssembly System Interface) [10] and (2) Lucet, a standalone Wasm compiler toolchain [22]. We discuss the impact of these below.

WASI is a Clang backend that produces Wasm binaries that do not require an embedding runtime (such as the Emscripten embedding runtime discussed in Section 2). Instead, the produced binaries use WASI-libc [9], an implementation of libc that is compatible with Wasm. Not having to provide an embedding runtime greatly simplifies the use of Wasm toolchains for library sandboxing.

Lucet is a standalone compiler built on top of the Cranelift code generator that can generate an ELF binary from a WASM module. While, Lucet eliminates the circuitous route of generating binaries we have relied on thus far, it currently has some limitations. At the time of writing, Lucet does not provide a convenient way to register and unregister callbacks dynamically [27]. Further, despite being a dedicated Wasm compiler, it still incurs significant overhead—using Lucet in the benchmark discussed in Section 5 incurs a 110% overhead.

Despite current limitations, the importance of the above tooling and specifications is well acknowledged. A recent effort called called the ByteCode Alliance [7] aims to bring together various tools and specifications such as Lucet, WASI, Cranelift, multiple Wasm runtimes under a single umbrella.

However, even with the availability of better tooling, many challenges in using Wasm based library sandboxing remain, including compiler modifications due to ABI differences (Section 3.1), linking and threading support (Section 4), performance limitations (Section 5.3), challenges in using hardware primitives to enforce Wasm guarantees (Section 6). We thus, reiterate our call to the Wasm and SFI research communities to work to improve Wasm module sandboxing as both communities can get significant benefits from high quality and performant sandboxed tooling.